\documentclass[10pt,a4paper,twocolumn]{article}
\usepackage[left=2.00cm, right=2.00cm, top=2.54cm, bottom=1.9cm]{geometry}

\usepackage{notation}

\title{\LARGE \bf
Exponential stochastic stabilization of a two-level quantum system via strict Lyapunov control
}

\author{Gerardo Cardona$^{1,2}$, Alain Sarlette$^{2,3}$ and Pierre Rouchon$^{1,2}$  
\thanks{$^{1}$Centre Automatique et Syst\`emes, Mines-ParisTech, PSL Research University. 60 Bd Saint-Michel, 75006 Paris, France.
        }%
\thanks{$^{2}$QUANTIC lab, INRIA Paris, rue Simone Iff 2, 75012 Paris, France
}
\thanks{$^{3}$Electronics and Information Systems Department, Ghent University, Belgium. Corresponding author,
	{\tt\small alain.sarlette@inria.fr}%
}
}
\date{}
\begin{document}

\maketitle
\thispagestyle{empty}
\pagestyle{empty}

\begin{abstract}
	This article provides a novel continuous-time state feedback control strategy to stabilize an eigenstate of the Hermitian measurement operator of a two-level quantum system. In open loop, such system converges stochastically to one of the eigenstates of the measurement operator. Previous work has proposed state feedback that destabilizes the undesired eigenstates and relies on a probabilistic analysis to prove convergence. In contrast, we here associate the state observer to an adaptive version of so-called Markovian feedback (essentially, proportional control) and we show that this leads to a global exponential convergence property with a strict Lyapunov function. Furthermore, besides the instantaneous measurement output, our controller only depends on the single coordinate along the measurement axis, which opens the way to replacing the full state observer by lower-complexity filters in the future.
\end{abstract}
\section{Introduction}
After the technological developments of the last decade, we are now at a stage where measuring and controlling quantum systems is experimentally realizable, as in the first real time quantum feedback experiment \cite{sayrin2011real}. From this pioneering work, involving essentially discrete-time logic, the physics community has been moving towards engineering designs that are truly promising for quantum computation \cite{campagne2016using,devoret2013superconducting,LiuShankarOfekEtAl2016} and which pose control problems in terms of continuous-time stochastic systems.

A basic element in quantum control is the use of quantum non-demolition (QND) measurements \cite{HarocheBook}. This is essentially a continuous-time version of the projection postulate, where performing a continuous measurement makes the quantum state progressively converge to a random eigenstate of the measurement operator. Each such eigenstate is a steady-state of the dynamics, and hence remains unperturbed under the backaction associated to this quantum measurement. QND eigenstates are thus natural equilibria of a measured quantum system, and stabilizing such a system towards one target eigenstate thanks to an appropriate feedback law would be a basic building block towards more involved control procedures. Several papers have indeed considered ways to stabilize a target QND eigenstate, see e.g.~\cite{van2005feedback,yamamoto2007feedback,mirrahimi2007stabilizing,tsumura2007global,tsumura2008global}. They prove global asymptotic convergence with probabilistic arguments, and based on a state feedback controller assuming that a state observer (often called the quantum Bayes filter) efficiently captures the evolution of the quantum state.

In our sense these results leave room for improvement in two directions. First, the feedback laws proposed in existing work are quite complicated, especially their implementation based on a quantum state observer would scale poorly with increasing system dimension as will be needed in advanced quantum technology. Second, the associated convergence analysis is rather involved and none of these feedback schemes are shown to ensure exponential convergence. It is well-known that exponential convergence is an indication of robustness when the system would interact with other subsystems or would perform subject to some perturbations. In other basic quantum settings, exponential convergence does hold with rather direct proofs. This is the case for the stabilization of target states that are not eigenstates of the measurement operator, via so-called Markovian feedback \cite{wiseman1994quantum,wiseman-milburnBook}. This essentially comes down to proportional control, and in \cite{wiseman2002bayesian} it was shown how the closed-loop equation can advantageously be reformulated as having changed the dissipation operators to have the target as steady state. A drawback of this approach is that, under non-ideal measurement conditions, a steady state far from the QND eigenstate would be subject to significant noise; a state close to a QND eigenstate in contrast would have little noise, but the Markovian feedback gain and the associated convergence rate towards such state approaches zero as the target approaches a QND eigenstate. Nonwithstanding, the convergence of the open-loop system under QND measurement towards \emph{the set} of its steady states can also be shown to be exponential. The absence of a proven similar property for the selection of one target QND steady state thus appears as an avoidable gap.

In the present paper, we identify a solution to the latter problem, i.e.~exponential stabilization of a QND eigenstate at least for a qubit system, and we pave the way towards addressing the first issue, i.e.~the real-time implementation complexity associated to a full state observer. Our solution in fact combines the state feedback ideas of \cite{van2005feedback,yamamoto2007feedback,mirrahimi2007stabilizing,tsumura2007global,tsumura2008global} with the Markovian feedback in the sense of Wiseman~\cite{wiseman1994quantum}. It should be noted that the idea of using this kind of dynamical feedback is also considered in~\cite{zhang2018locally,MoelmerChernNumber}, but for different goals. As far as we know our proposal is the first one that provides a convergence proof ensuring almost-sure exponential asymptotic stability with a strict Lyapunov function in the stochastic sense. The interpretation of the control law comes down to an adaptive version of the proportional feedback gains, depending on the distance of the measured QND coordinate from its target value. This coordinate is still assumed to be estimated by a perfectly converging quantum filter for now, but in the future it seems natural to replace it by lower-order filters.

The structure of the paper is as follows. In section \ref{section:1} we set up some preliminaries on quantum systems under continuous measurements in finite dimensions. We recall the asymptotic behavior for the open loop system and provide an exponential convergence rate for a general QND system, via an original Lyapunov function, under a non-degeneracy condition. In section \ref{section:3} we review static output feedback (Markovian feedback). We explain how to tune the feedback gain as a function of the target steady state. We provide a Lyapunov function that proves the exponential convergence towards any chosen target, except that the convergence rate tends to $0$ for a QND eigenstate as target. This highlights the necessity of another approach for QND eigenstates. In section \ref{section:4} we consider state feedback control, we review some of the existing results on global stabilization for QND eigenstates and discuss the main difficulties related to this kind of approach. In section \ref{section:5} we present our main result, i.e.~the novel state feedback controller and its exponential convergence proof.

\section{QND diffusive  measurements}\label{section:1}

We consider the basic  open-loop model  for a quantum system under Quantum Non Demolition (QND) continuous measurements \cite{barchielli2009quantum} in finite dimension $1<n<\infty$ :
\begin{align}\label{eq:SMEQND}
d\rho&=\damping(L,\rho)dt+\sqrt{\eta} \measurement(L,\rho)dW,\\
dY&=\sqrt \eta \tr(L\rho+ \rho L^\dag)dt+dW,
	\label{eq:SMEQNDObs}
\end{align}
where $\rho$ belongs to the set of density matrices $\mathcal{S}=\{\rho\in \C^{n\times n} : \rho\ge 0,\ \rho=\rho^\dagger, \ \tr (\rho )=1 \}$, $\tr(\ \cdot \ )$ denotes the trace, $W$ is a standard Wiener process and $dY$ is the measurement process, $L=L^\dag$ is a Hermitian matrix  in  $\C^{n\times n}$. The general properties of  such stochastic differential equations~\eqref{eq:SMEQND} are well known:  existence and uniqueness of solutions in the  state-space $\mathcal{S}$. We have used  the following usual notations:
\begin{align}
   \damping(L,\rho)&=L\rho L^\dagger -\frac{1}{2}L^\dagger L \rho -\frac{1}{2} \rho L^\dagger L ,\\
   \measurement(L,\rho)&=L\rho +\rho L^\dagger -\tr((L+L^\dagger )\rho)\rho.
\end{align}
Throughout the paper we assume that $L=L^\dag$ admits a non degenerate spectrum: all its eigenvalues $(\lambda_{\ell})_{1\leq \ell \leq n}$ are distinct and its normalized eigenvectors  $(\ket{\bar\psi_{\ell}})_{1\leq \ell  \leq n}$, with the bra-ket notation (see, e.g.,  \cite{nielsen2002quantum}), form an orthonormal frame.

Such QND systems behave in a particular way:  each eigenvector of $L$, thus $\bar{\rho}_{\ell} \triangleq \ketbra{\bar{\psi}_{\ell}}{\bar{\psi}_{\ell}}$, is a steady-state of Eq. \eqref{eq:SMEQND};  using It\^o calculus and  taking the expectation we have
\begin{equation}
\frac{d}{dt}\Exp(\tr(\rho\bar{\rho}_{\ell} ))=0,
\end{equation}
where $\tr(\rho\bar{\rho}_{\ell} )= \bra{\bar{\psi}_{\ell}} \rho \ket{\bar{\psi}_{\ell}}$ corresponds to the so-called trace fidelity, reaching its maximum $1$ only when $\rho=\bar{\rho}_{\ell}$.  Thus $\tr(\rho\bar{\rho}_{\ell} )$ is a martingale for~\eqref{eq:SMEQND}. Moreover as shown in~\cite{van2005feedback,mirrahimi2007stabilizing,BenoiP2014}, all trajectories converge to one of the above steady states, with $\tr(\rho_t\bar{\rho}_{\ell} )$ giving the probability to end up on $\bar{\rho}_{\ell}$ conditioned on knowing $\rho_t$ (in particular, for $t=0$).

We propose in the following lemma a proof of this asymptotic result where we estimate the convergence rate via an original exponential Lyapunov function.
\begin{Lemma}[Exponential stability of QND systems]\label{prop:OpenLoop}
	Consider equation \eqref{eq:SMEQND} with $L$ a non-degenerate Hermitian matrix and denote as $\bar{\rho}_{\ell}$ the states $\{\bar{\rho}_{\ell}:= \ketbra{\bar{\psi}_{\ell}}{\bar{\psi}_{\ell}} : L\ket{\bar{\psi_{\ell}}}=\lambda_{\ell}\ket{\bar{\psi_{\ell}}} , \;\; 1\le \ell \le n\}$. 	
Let
$$
V(\rho) =\sum_{\ell'=1}^n \sum_{\ell<\ell'}\sqrt{\tr(\rho\bar \rho_{\ell}) \tr(\rho\bar \rho_{\ell'}) }
.
$$
Then
$$\Exp (V(\rho_t)~|~\rho_0) \le e^{-rt}V(\rho_0)$$
for all $t\geq 0$,  with rate $r\geq 0$ given by $$
			r=\tfrac{1}{2}\eta\;\min_{\ell',\ell\neq\ell'} \left((\lambda_{\ell}-\lambda_{\ell'})^2\right ).
$$    	
\end{Lemma}
\begin{proof}
Let $\xi_{\ell}:=\sqrt{\tr(\rho\bar{\rho}_{\ell})}$ for $1\le \ell \le n$. It satisfies
\begin{equation}\label{eq:dxiell}
	\quad d\xi_{\ell}=-\tfrac{1}{2}\eta(\lambda_{\ell}-\varpi(\xi)))^2 \xi_\ell dt +  \sqrt{\eta}(\lambda_{\ell}-\varpi(\xi)))\xi_{\ell}dW
\end{equation}
with $\xi=(\xi_{\ell'})_{1\leq \ell'\leq n}$ and $\varpi(\xi)=\sum_{\ell'} \lambda_{\ell'} \xi_{\ell'}^2$. With the $\xi$ coordinates, $V$ becomes a $C^2$ function of $\xi$, $V(\xi)=\sum_{\ell',\,\ell < \ell'} \xi_{\ell}\xi_{\ell'}$,  with $\xi$ obeying the system of $n$ stochastic differential equations \eqref{eq:dxiell} driven by a common scalar Wiener process $W$. We can directly compute the Markov generator $\mathcal{A}$ of the $\xi$-system on $V$ using the formula~\eqref{App:DiffOp} recalled in appendix.  Standard algebraic computations yield
$$
\mathcal{A} V =- \tfrac{\eta}{2} \sum_{\ell'=1}^n \sum_{\ell<\ell'}   ( \lambda_{\ell}-\lambda_{\ell'} )^2  \;\xi_{\ell} \xi_{\ell'}.
$$
Since each component of $\xi(t)$ remains non-negative for all $t$, we have
$
\mathcal{A} V  \leq -   \tfrac{\eta}{2} \left(\min_{\ell',\ell\neq\ell'} (   \lambda_{\ell}-\lambda_{\ell'})^2 \right) V
$ and we conclude with Gronwall's inequality.
\end{proof}

Since $V(\rho)=0$ is equivalent to $\{\rho=\bar \rho_{\ell} \text{ for some } \ell\}$, standard results on stochastic stability imply that, for any initial state $\rho_0\in\mathcal {S}$, the solution $\rho_t$ of~\eqref{eq:SMEQND} converges almost surely to one of the states $\bar\rho_{\ell}$, as proved independently in \cite{van2005feedback,mirrahimi2007stabilizing}.

As explained in~\cite{stockton2004deterministic}, such QND diffusive measurement can be viewed as preparing a pure state among the $\ket{\bar{\psi}_{\ell}}$ in a non-deterministic manner. As recalled in our introduction, several papers have proposed feedback controllers to actually stabilize one target steady state (see e.g.~\cite{van2005feedback,yamamoto2007feedback,mirrahimi2007stabilizing,tsumura2007global}), establishing global convergence with probabilistic arguments. In section \ref{section:5} we propose a new controller, based on adapting so-called quantum Markovian feedback \cite{wiseman1994quantum}, for which we can prove exponential convergence with a Lyapunov function like in the open-loop case. In the next section we therefore first recall the stabilization properties of such Markovian feedback, or static output feedback, which are in fact complementary to our goal.


\section{Static output feedback for QND systems }\label{section:3}

When adding a control input to the system, we have:
\begin{multline} \label{eq:QNDu}
    d\rho= \exp(-iF\, u dt)\, \left(\rho + \damping(L,\rho)dt+\sqrt{\eta}\measurement(L,\rho)dW \right)\\ \,\exp(iF\, u dt) \quad -\rho \; .
\end{multline}
Here $u$ is a (real) control signal and $F$ is a Hermitian matrix. We use the propagator notation to emphasize that when $u\,dt$ contains Wiener processes, as we will do, $\exp(-iF\, u dt)$ must be expanded to second order in a causal It\={o} calculus. In analogy to classical systems,  consider the simplest control scheme, a proportional output feedback of the form
\begin{equation}\label{eq:PropControl}
udt= f  dt+ \kappa dY,
\end{equation}
where $\kappa$ is a constant  gain and $f$ a constant bias.  

Following~\cite{wiseman1994quantum} and It\=o rules, the static controller \eqref{eq:PropControl} leads to the following closed-loop stochastic master equation:
\begin{multline}\label{eq:SFME}
d\rho =-if \comm{F}{\rho}dt-\tfrac{i \kappa \sqrt{\eta}}{2}\comm{ FL+L F}{\rho}dt\\ +\damping(L-i\kappa  \sqrt{\eta}F,\rho)dt+\damping(i\kappa  \sqrt{1-\eta}F,\rho)dt\\+\sqrt{\eta}\measurement(L,\rho)dW-i\kappa\comm{F}{\rho}dW.
\end{multline}
The next lemma shows that the controller \eqref{eq:PropControl} cannot be tuned to achieve global asymptotic stability towards any QND eigenstate $\bar{\rho}_\ell$.

\begin{Lemma}\label{lem:outputfeedback1}
There exists no combination of  $F$, and constants $f$ and $\kappa$ such that the proportional output feedback~\eqref{eq:PropControl} would yield a closed-loop dynamics~\eqref{eq:SFME} which globally converges towards a target  QND eigenstate $\bar{\rho}_\ell$ for some chosen $\ell\in\{1,\ldots,n\}$.
\end{Lemma}
\begin{proof}
 We can rewrite Eq. ~\eqref{eq:SFME} as	
\begin{multline*}
d\rho =-if \comm{F}{\rho}dt-i \kappa \sqrt{\eta}\comm{ F}{L\rho+\rho L}dt\\ +\damping(L,\rho)dt+\kappa^2\damping(F,\rho)dt\\+\sqrt{\eta}\measurement(L,\rho)dW-i\kappa\comm{F}{\rho}dW.
\end{multline*}	
  Then if $\bar{\rho}_\ell$ is a steady-state of~\eqref{eq:SFME} and since $f,\kappa$ are constants, then $-if \comm{F}{ \bar{\rho}_\ell}-i \kappa \sqrt{\eta}\comm{ F}{L\bar{\rho}_\ell+\bar{\rho}_\ell L}+\damping(L,\bar{\rho}_\ell)+\kappa^2\damping(F,\bar{\rho}_\ell)=0$  and $\sqrt{\eta}\measurement(L,\bar{\rho}_\ell)-i\kappa \comm{F}{\bar{\rho}_\ell}=0$. Since $L \bar{\rho}_\ell = \lambda_\ell \bar{\rho}_\ell$, the second condition implies that  $\kappa [F,\bar{\rho}_\ell]=0$. By plugging this into the other steady-state constraint, we get $f \comm{F}{ \bar{\rho}_\ell}=0$ and $\kappa^2\damping(F,\bar{\rho}_\ell)=-\tfrac{\kappa^2}{2}\comm{F}{\comm{F}{\bar{\rho}_\ell}}=0$. If $ f=\kappa=0$, we are in open-loop and $\bar{\rho}_\ell$ is not globally asymptotically stable. When $\kappa$ or $f$ are not zero, we must have $\comm{F}{ \bar{\rho}_\ell}=0$. But then we get $d\left(\tr(\rho_t  \bar{\rho}_\ell) \right)=0$, thus $\tr(\rho_t  \bar{\rho}_\ell) = \tr(\rho_0  \bar{\rho}_\ell)$ is time invariant  and we cannot have global convergence towards $\bar{\rho}_\ell$.
\end{proof}

This obstruction seems to only appear for QND eigenstates $\bar{\rho}_\ell$. The next lemma indeed shows that, at least for a two-level system, any other pure state can be exponentially stabilized by a static output feedback with a fixed measurement operator $L$ and detection efficiency $\eta=1$. We use the standard Pauli matrix and Bloch sphere notation for the qubit system.
\begin{Lemma}\label{lem:PropCtrlSynth}
	Consider~\eqref{eq:SFME} with $n=2$, $\eta=1$, and
     $L=\sqrt{\tfrac{\Gamma}{2}}\sigma_z=\sqrt{\tfrac{\Gamma}{2}}\begin{bmatrix}1 & 0\\ 0 &- 1 \end{bmatrix},$
    $F=\sigma_y=\begin{bmatrix}0 & -i\\ i &0 \end{bmatrix}.$ Take $\bar{\theta}\notin \{ k\pi: k \in \mathbb{Z} \}$ and set  $f=-\tfrac{\Gamma}{2}\sin(\bar\theta)\cos(\bar{\theta})$, $\kappa=\sqrt{\tfrac{\Gamma}{2}}\sin(\bar{\theta})$. Then the closed-loop system exponentially stabilizes the pure state:
	$$
	\bar \rho=\frac{1}{2}\begin{bmatrix}
	1+\cos(\bar{\theta})&\sin(\bar{\theta})\\ \sin(\bar{\theta})& 1-\cos(\bar{\theta})
	\end{bmatrix}	,
	$$
as the Lyapunov function $V(\rho) =1-\tr(\rho \bar\rho)$ decreases according to $\Exp (V(\rho_t)) = e^{-rt}V(\rho_0)\ \forall t\ge 0$ with rate $r= \Gamma(\sin\bar\theta)^2$.
\end{Lemma}
The proof, left to the reader, is based on  standard matrix manipulations showing 	$ d\Exp(V) = -\Gamma(\sin\bar{\theta})^2 \Exp (V) ~dt$.  By modifying the actuation Hamiltonian as $F=U_{\bar\alpha} \sigma_y U_{\bar\alpha}^\dag$, where $U_{\bar\alpha}= \exp(-i\bar \alpha \sigma_z/2)$ is a rotation of angle $\bar\alpha$ around the $z$ axes of the Bloch sphere associated to the qubit, the same feedback gains stabilize $U_{\bar\alpha} \bar \rho U_{\bar\alpha}^\dag$. With $\bar\alpha\in[0,2\pi]$ and $\bar\theta \in]0,\pi[$, any pure state different from the two eigenstates of $\sigma_z$ are thus obtained via $U_{\bar\alpha} \bar \rho U _{\bar\alpha}^\dag$.

For more complex systems, with $n > 2$, we can use the algebraic criterion elaborated in \cite{ticozzi2008quantum,ticozzi2009analysis,ticozzi2010stabilizing} to analyze global convergence towards a unique pure state under such static output feedback. Thanks to positivity of $\rho$, this can be done by looking at the ensemble average closed-loop dynamics
$$
d\rho/dt  = -i f \comm{F}{\rho}-\tfrac{i \kappa}{2}\comm{ FL+L F}{\rho} +\damping(L-i\kappa F,\rho)
$$
and its convergence towards  pure states. We think that at least when the goal pure state $\bar \rho$ is such  that $\tr(\bar\rho \bar\rho_\ell)>0$ for all $\ell\in\{1, \ldots, n\}$, then a control Hamiltonian $F$ can be chosen among a ``reasonable'' set and $f$ and $\kappa$ exist which ensure global exponential convergence of~\eqref{eq:SFME} towards $\bar \rho$. At this point we leave the question of feedback constrained to typical available $F$ for $n>2$, and we move on to the stabilization of the elusive targets, that is the QND eigenstates $\bar\rho_\ell$.

\section{Quantum state feedback for QND systems}\label{section:4}

Stabilization by quantum state feedback of QND eigenstates has been explored and solved in several papers \cite{van2005feedback,yamamoto2007feedback,mirrahimi2007stabilizing,tsumura2007global}, with 
\begin{equation}\label{etroui}
u\, dt = f(\rho) \, dt
\end{equation}
 and no direct feedback of $dY$. Although they all start with Lyapunov control techniques and the open-loop martingale  $V(\rho)=1-\tr(\rho \bar\rho_\ell)$, they also involve specific stochastic arguments combining support of the closed-loop trajectories with Doob inequalities in order to establish the global asymptotic almost sure convergence for the closed-loop system \cite{mirrahimi2007stabilizing,tsumura2007global}. Let us explain the basic reason why it is difficult, as already highlighted in~\cite{van2005feedback}, to construct quantum state feedback and a strict closed-loop Lyapunov function in this case.

With~\eqref{eq:QNDu},\eqref{etroui}, we have $\Exp (V_{t+dt}| \rho_t,u_t)= V_t + u_t\tr(i\comm{F}{\rho_t}\bar\rho_\ell) dt$. The quantum state feedback $u_t=-\tr(i\comm{F}{\rho_t}\bar\rho_\ell)$ makes this expectation decreasing. However any QND state $\bar\rho_{\ell'}$ with $\ell'\neq \ell$ is a closed-loop steady state since  $ \tr(\comm{F}{\bar\rho_{\ell'}}\bar\rho_\ell)=0$, for any choice of $F$. In a stochastic convergence setting, such steady states are harder to treat, as significant portions of the state can even converge to maxima of a stochastic Lyapunov function. Researchers have proposed several ways to solve this problem.

In all pure Lyapunov  feedback approaches like \cite{van2005feedback,yamamoto2007feedback}, the presence of several steady states remains, and an additional proof element e.g.~based on sending undesired equilibria to infinite values of the Lyapunov function is used to obtain almost global stability. In \cite{mirrahimi2007stabilizing} and in \cite{tsumura2007global}, the controllers are explicitly designed to apply perturbations which remove the undesired steady states. In \cite{mirrahimi2007stabilizing} this is done with a discontinuous control that gives a constant input whenever $\rho$ is too close to the bad QND eigenstates $\bar\rho_{\ell'}$ and then switching to the above Lyapunov feedback once $\rho$ is close enough to the goal state $\bar\rho_\ell$. In~\cite{tsumura2007global} a continuous approach is proposed where the control is perturbed by the addition of a term that only vanishes on the target state. More precisely, for the two-level system of lemma~\ref{lem:PropCtrlSynth}, a smooth feedback law
\begin{equation}\label{eq:TsumuEx}
    u=-\alpha\tr(i\comm{\sigma_y}{\rho}\sigma_z) + \beta(1-\tr(\rho \sigma_z)),
\end{equation}
with $\alpha,\beta>0$ such that $\frac{\beta^2}{8\alpha\eta }<1$, is applied to stabilize the excited state $\bar\rho_1=(1+\sigma_z)/2$. The convergence proof then involves a combination of Lyapunov function on part of the state space, and probabilistic arguments on the other part. With the feedback \eqref{eq:TsumuEx}, the set $\{\rho\in \mathcal{S}:V(\rho) \leq \frac{\beta^2}{8\alpha\eta }  \}$ satisfies $ d\Exp (V)\leq 0$ defining a kind of attraction set for the control Lyapunov function. Thus trajectories that do not exit this region of attraction will converge to the target. Outside the region of attraction, both feedback schemes rely on ensuring that trajectories will reach again this region of attraction. We refer to \cite{mirrahimi2007stabilizing,tsumura2007global,tsumura2008global} for more details.

All those methods only establish asymptotic convergence. Having a stronger global convergence result like exponential convergence is important for robustness and estimation of convergence speed. The robustness is practically important in particular towards unmodeled dynamics. This could concern other system elements like actuators, but at very least there is the quantum filter which estimates the state $\rho$ from the measurement inputs with a finite convergence speed. The convergence speed is particularly important in applications where feedback control is applied to protect the fragile quantum systems from perturbing effects, like decoherence (see simulations section). While our paper was under review, a parallel paper~\cite{HadisCDC2018} has proposed an adaptation of the feedback and of several stochastic techniques used in \cite{tsumura2007global}. They obtain global asymptotic convergence, which is further proved to be asymptotically exponential, but without giving guarantees at finite time. As far as we know, a feedback scheme ensuring global exponential convergence via simple Lyapunov argument has remained an open issue.

\section{Exponential feedback stabilization of two-level  QND systems}\label{section:5}

We present now our main result. We consider \eqref{eq:QNDu} with $L=\sqrt{\tfrac{\Gamma}{2}}\sigma_z$, $F=\sigma_y$, where $\sigma_z,\sigma_y$ are as in Lemma \ref{lem:PropCtrlSynth}.
When $u=0$, Lemma~\ref{prop:OpenLoop} shows that $\rho_t$ converges  as $t\rightarrow\infty$ to the set of states $\{\ketbra{e}{e},\ketbra{g}{g}\}$  corresponding to the eigenvalues $1$ and $-1$ respectively of $\sigma_z$. We propose a feedback law combining quantum-state  and  output feedback, in the form:
\begin{equation}\label{eq:FeedbackLaw}
u_tdt=f(\rho_t)dt+\kappa(\rho_t)dY
\end{equation}
where $f$ and $\kappa$  are regular functions of $\rho$ such that the closed-loop dynamics
\begin{multline}\label{eq:SMEGlobal}
     d\rho_t =-if(\rho)\comm{\sigma_y}{\rho}dt+\damping\left(\sqrt{\tfrac{\Gamma}{2}}\sigma_z-i\sqrt{\eta} \kappa(\rho) \sigma_y,\rho\right)dt
        \\+\damping\left(i\sqrt{(1-\eta)} \kappa(\rho) \sigma_y,\rho\right)dt
     \\+\sqrt{\tfrac{\eta \Gamma}{2}}\measurement(\sigma_z,\rho)dW -i\kappa(\rho) \comm{\sigma_y}{\rho}dW
\end{multline}
is  well-posed. For any realization $\rho_t$, if $\rho_0 $ belongs to the compact set $\mathcal{S}$ of density operators, then $\rho_t$ remains in $\mathcal{S}$  for all $t>0$. We will target the stabilization of the state $\ketbra{e}{e}$; the treatment for target $\ketbra{g}{g}$ is exactly similar. Theorem~\ref{thm:StateControl} below shows that $f$ and $\kappa$ can be chosen  in order to have  an explicit strict closed-loop Lyapunov function, converging in average exponentially to zero, and implying the global convergence of the state towards $\ketbra{e}{e}$.
\begin{Theorem}\label{thm:StateControl}
  Consider equation \eqref{eq:SMEGlobal} with
  $$
  f(\rho)=-\tfrac{\eta\Gamma}{2}(1-\tr(\rho\sigma_z)^2)  \text{ and } \kappa(\rho)=\sqrt{ \tfrac{\eta\Gamma}{2}}(1-\tr(\rho\sigma_z)).
  $$
  Set $ V(\rho)=\sqrt{1-\tr(\rho\sigma_z)}$.

  Then $\ketbra{e}{e}$ is globally exponentially stable (in the sense of definition \ref{def:stab} in appendix with $p=1$) and  $\Exp( V (\rho_t)~|~\rho_0) \le e^{-rt}V(\rho_0)$ for all $ t \ge 0$ and $\rho_0\in\mathcal{S}$,  with convergence rate $ r = \frac{\eta \Gamma }{4}.$
\end{Theorem}

\begin{proof}
	Let $z=\tr(\rho\sigma_z)$,  $x=\tr(\rho\sigma_x)$,  $y=\tr(\rho\sigma_y)$ and set  $ \xi=\sqrt{1-z}$. We express $f=-\tfrac{\eta \Gamma}{2}(2-\xi^2)\xi^2$, $\kappa=\sqrt{\tfrac{\eta \Gamma}{2}}\xi^2 $.

After the change of coordinates $(x,y,z)\mapsto(x,y,\xi)$, system \eqref{eq:SMEGlobal} satisfies:
\begin{align*}
  dx=&\left( \eta \Gamma(2\text{-}\xi^2)\xi^2(1\text{-}\xi^2)+2\eta \Gamma \xi^2 -\eta \Gamma\xi^4x \right)dt\\
	&+\sqrt{2\Gamma \eta}((1-\xi^2)(\xi^2 -x))dW,\\
	dy=&-\Gamma ydt-\sqrt{2\Gamma \eta}(1-\xi^2)ydW,\\
	d\xi=&\tfrac{\eta\Gamma}{2}\left( (2-\xi^2)x -\xi^2 (1-\xi^2) -\tfrac{1}{2}((2-\xi^2)-x)^2 \right)\xi dt\\
	&-\sqrt{\tfrac{\eta\Gamma}{2}}((2-\xi^2)-x)\xi dW.
\end{align*}
We now compute the Markov generator $\mathcal{A}$ of the Lyapunov function $V(x,y,\xi)=\xi$ which yields
\begin{multline*}
 \mathcal{A}V= \tfrac{\eta\Gamma}{2}\left ( -\xi^2 (1-\xi^2)-\tfrac{1}{2}((2-\xi^2)^2+x^2)\right)\xi\\
=   -\tfrac{\eta \Gamma}{2}(\tfrac{1}{2}+\tfrac{3}{2}z^2+ \tfrac{1}{2}x^2 )\xi\\
            \leq  -\tfrac{\eta \Gamma}{4}\xi.
\end{multline*}
%
Since $ d\mathbb{E}(V_t)= \Exp(\mathcal{A} V(\xi)) dt$,  by Gronwall's inequality $\Exp (V_t~|~V_0)\le e^{-rt}V_0$.

To establish convergence of the state, we use theorem~\ref{appendix:Stability} in the appendix. The Lyapunov function $V$ is non-negative and it is a $C^2$ function of $(x,y,\xi)$. It is obviously upper bounded by $V = \xi < k_2 (\sqrt{x^2+ y^2 + \xi^2})^p$ with $k_2 = p=1$. For a lower bound, we must take into account that the system evolves in the set $\mathcal{S}=\{x,y,\xi : x^2+y^2+(1-\xi^2)^2\leq 1 \}$; a direct development yields $x^2 + y^2 \leq \xi^2(2-\xi^2) \leq 2 \xi^2$, such that $V = \xi \geq (\sqrt{(x^2+ y^2 + \xi^2)/3})^p = k_1 (\sqrt{x^2+ y^2 + \xi^2})^p$ with $p=1$ and $k_1 = 1/\sqrt{3}$. The latter also implies that $\mathcal{A} V(\xi) \leq - k_3 (\sqrt{x^2+ y^2 + \xi^2})^p$ with $p=1$ and $k_3 = r k_1$. Thus all the assumptions of theorem~\ref{appendix:Stability} are satisfied and we conclude that $(x,y,\xi)=(0,0,0)$ is exponentially stable in the sense of definition \ref{def:stab} in appendix with $p=1$.
\end{proof}

With Lemma \ref{lem:PropCtrlSynth} and Theorem \ref{thm:StateControl} we have achieved full exponential stabilization of a two-level system. The static output feedback of Lemma \ref{lem:PropCtrlSynth} can stabilize all states of a qubit except a QND eigenstate; the feedback of Theorem \ref{thm:StateControl} stabilizes the QND eigenstates. The overhead cost in Theorem \ref{thm:StateControl} with respect to static output feedback is that an estimator is needed for the state component $\tr(\rho\sigma_z)$. Interestingly, this is much less than the full quantum state $\rho$, and it corresponds to the component $\tr(L\rho+ \rho L^\dag)dt$ that is directly present in the measurement output $dY$ with $L\propto \sigma_z$. This shows that besides allowing a more standard exponential convergence proof, our controller also possibly opens the door to simpler feedback schemes than full-state-feedback.

Regarding convergence speed, the result of Theorem \ref{thm:StateControl} is at most $4$ times slower than the open loop convergence towards a random eigenstate (Lemma \ref{prop:OpenLoop}). This is much faster than the slow convergence of Lemma \ref{lem:PropCtrlSynth} for target states close to $\ket{e}\bra{e}$, and one may envision to speed up the latter by adapting the strategy of Theorem \ref{thm:StateControl}.

\subsection*{Numerical test}

\begin{figure}[h]
	\centering
	\includegraphics[width=\linewidth]{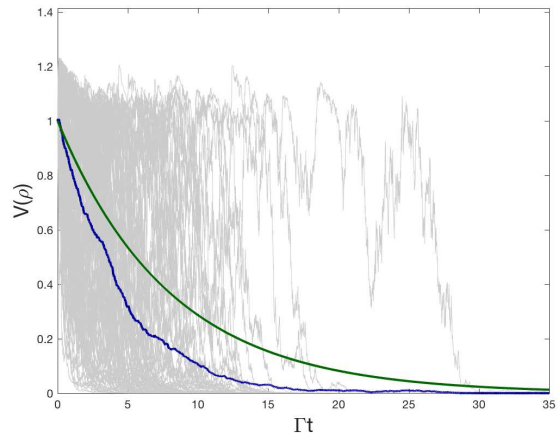}
	\caption{ In blue: Ensemble average of $V(\rho)=\sqrt{1-\tr(\rho\sigma_z)}$ over 100 trajectories (in gray) of the closed-loop system \eqref{eq:SMEGlobal} with initial condition $\rho_0=I/2$. In green: Bound obtained from theorem \ref{thm:StateControl}.}
	\label{fig:perfectmeasurements}
\end{figure}

\begin{figure}[h]
	\centering
	\includegraphics[width=\linewidth]{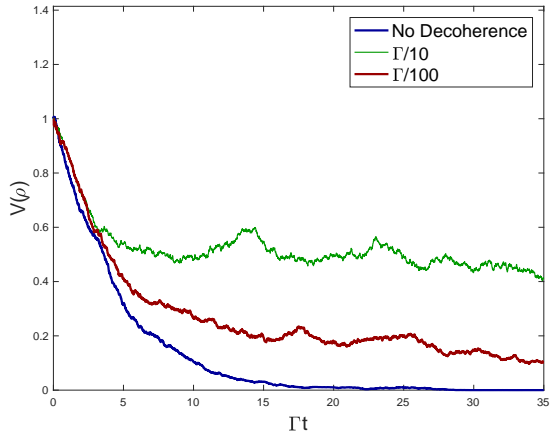}
	\caption{Comparison of the ensemble average of $V(\rho)=\sqrt{1-\tr(\rho\sigma_z)}$ over 100 trajectories when adding to \eqref{eq:SMEGlobal} a decoherence term $\tfrac {\tilde\Gamma}{2}\damping(\ketbra{g}{e},\rho)$ with $\tilde\Gamma \in \{0,\Gamma/100,\Gamma/10\}$.} 
	\label{fig:coherences}
\end{figure}

We test the results of Theorem \ref{thm:StateControl} through numerical simulations. We fix $\eta=0.5$ and consider 100 trajectories to sample the stochastic behavior. In fig. \ref{fig:perfectmeasurements} we show the behavior of the closed loop system \eqref{eq:SMEGlobal} with initial condition $\rho_0=I/2$. The simulation appears to confirm the exponential convergence. In fig. \ref{fig:coherences}, we show the behavior when we add a standard perturbation to the nominal behavior, namely a decoherence term of the form $\tfrac {\tilde\Gamma }{2}\damping(\ketbra{g}{e},\rho)$ that expresses energy loss of the excited state $\ket{e}\bra{e}$ by spontaneous photon emission. By varying $\tilde \Gamma \in \{0,\Gamma/100,\Gamma/10\}$, we see that the Lyapunov function goes from about $0$ to about $0.1$ and $0.5$, which corresponds to $z=1$, $z=.99$ and $z=.75$ respectively.

\section{Conclusion}
A quantum system under quantum non-destructive (QND) measurements is a particular stochastic system, where any eigenstate of the measurement operator is a steady state of the stochastic dynamics. While stabilizing a particular eigenstate under continuous-time QND measurement appears as a basic problem of quantum control, it is also remarkably different from standard control problems. Indeed, we have first considered a proportional output feedback (see Lemma \ref{lem:PropCtrlSynth}), with which we recall that we can stabilize all target states except the QND steady states. This indicates that for the QND eigenstates as targets, one must make a more involved controller in order to bias the stochastic evolution towards the desired extremum. Previous work like \cite{van2005feedback,yamamoto2007feedback,mirrahimi2007stabilizing,tsumura2007global} has turned to specific stochastic controller designs to achieve this goal, on the basis of a quantum state observer. In this paper we have shown that just adapting the proportional output feedback gain, as a function of estimated quantum state, is sufficient to obtain global exponential stabilization, something that previous state feedback controls do not achieve or at least could not prove. In our case, the proof is a direct consequence of standard stochastic convergence theorems with a strict Lyapunov function. Interestingly, this controller architecture can be re-interpreted as indirectly engineering state-dependent dissipation operators.

While this provides a simple and well-proved state feedback law, the complexity of its verbatim implementation would still be dominated by the quantum state observer, which is needed to estimate the quantum state on the basis of the measurement outputs. However, our controller depends on the qubit state through only the $z$ coordinate,  which is the \emph{measured} coordinate. This paves the way towards using simpler filters to extract accurate $z$ information from the measurement output, and leading to more scalable controllers. The extension of this technique to higher-dimensional systems is indeed a subject of ongoing work, with promising preliminary results.




\bibliographystyle{IEEEtran}
\bibliography{IEEEabrv,Biblio}


\section{Appendix}


We recall some results from stochastic stability and  refer the reader to \cite{kushner1972stochastic,Kushner1967v,khasminskii2011stochastic} for the proof of these results and for further reference.
We consider diffusion processes $x_t$ on $\R^n$. They  correspond to  solution of It\={o} stochastic differential equations of the form \begin{equation}\label{eq:SDE}
dx=\mu(x)dt+\sigma(x)dW,
\end{equation}
where the $\mu,\sigma$ are regular functions of $x$ that satisfy the usual conditions for existence and uniqueness of SDE's \cite{arnold1974stochastic}.   For a $C^2$ function $V$ the Markov generator associated with \eqref{eq:SDE} is
\begin{equation}\label{App:DiffOp}
\mathcal{A}V=\sum_i \mu_i\frac{\partial}{\partial x_i}V+\frac{1}{2} \sum_{i,j} \sigma_i\sigma_j\frac{\partial^2}{\partial x_ix_j}V,
\end{equation}
and
$$
  \Exp(V(x_t)\ \lvert V(x_0) )=V(x_0)+\Exp\left (\int_0^t \mathcal{A}V(x_s)ds\right).
$$
In this appendix we assume that    $\mathcal{S}$, a compact subset of $\R^n$,  is positively  invariant. We use the following specific asymptotic stability definition.
\begin{Definition}[Khasminskii \cite{khasminskii2011stochastic}]\label{def:stab}
Consider the  diffusion process on $\mathcal{S}$ governed by~\eqref{eq:SDE} with  $0\in\mathcal{S}$, $\mu(0)=\sigma(0)=0$ and $p>0$. Consider the Euclidean norm on $\R^n$:  $\| x \| = \sqrt{\sum_i x_i^2}$.
 The equilibrium solution $x_t=0$  is said \textit{exponentially p-stable} if, for some constants $C>0$ and $r>0$ such that  $$ \Exp(\| x_t \|^p ~|~x_0)\le C \| x_0\|^p e^{-rt}.$$
 \end{Definition}

%

The stochastic counterpart of Lyapunov's second method provides a sufficient criterion for proving such asymptotic stability in terms of the expectation of a scalar function $V$ as shown by the following theorem.
\begin{Theorem}[Khasminskii \cite{khasminskii2011stochastic}]\label{appendix:Stability}
	
	Let $V(x)$ be a nonnegative real-valued twice continuously differentiable function with respect to $x\in \mathcal{S}$ everywhere except possibly at $x=0$. Assume that  exist strictly positive constants $p$, $k_1$, $k_2$ and $k_3$ such that for all $x \in \mathcal{S}$ we have
	\begin{enumerate}
			\item $\mathcal{A}V(x)\leq -k_3\| x \|^p $,
			\item  $k_1\| x \|^p \le V(x)\le k_2 \| x \|^p$.
	\end{enumerate}
	Then the equilibrium solution $x_t=0$ is {\em exponentially $p$-stable} in the sense of definition~\ref{def:stab}.



\end{Theorem}

We note that exponential convergence on a compact set $\mathcal{S}$ is stronger than almost sure convergence, and thus of all other weaker notions of convergence. This follows from dominated convergence argument to interchange the order of the limit and the expectation, see e.g.~\cite[Lemma 4.9]{mirrahimi2007stabilizing} for technical details.

\end{document}